\documentclass[final,3p,times,authoryear]{elsarticle}

\usepackage{graphicx}

\usepackage{amssymb}

\journal{Icarus}

\newcommand{\beq}{\begin{equation}}
\newcommand{\eeq}{\end{equation}}

\begin{document}

\begin{frontmatter}



\title{Extreme Sensitivity of the YORP Effect to Small-Scale Topography}


\author{Thomas S.~Statler}
\ead{statler@ohio.edu}
\address{Astrophysical Institute, Department of Physics and Astronomy,
251B Clippinger Research Laboratories, Ohio University, Athens, OH
45701, USA}

\begin{abstract}

Radiation recoil (YORP) torques are shown to be extremely sensitive to
small-scale surface topography, using numerical simulations. Starting from
a set of ``base objects'' representative of the near-Earth object population,
random realizations of three types of small-scale topography are added:
Gaussian surface fluctuations, craters, and boulders. For each, the expected
relative errors in the spin and obliquity components of the YORP torque caused
by the observationally unresolved small-scale topography are computed.
Gaussian power, at angular scales below an observational limit, produces
expected errors of order 100\% if observations constrain the surface
to a spherical harmonic order $l \lesssim 10$. For errors under 10\%, the
surface must be constrained to at least $l = 20$. A single crater with
diameter roughly half the object's mean radius, placed at random locations,
results in expected errors of several tens of percent. The errors scale with
crater diameter $D$ as $D^2$ for $D>0.3$ and as $D^3$ for $D<0.3$ mean radii.
Objects that are identical except for the location of a single large crater
can differ by factors of several in YORP torque, while being photometrically
indistinguishable at the level of hundredths of a magnitude. Boulders placed
randomly on identical base objects create torque errors roughly 3 times
larger than do craters of the same diameter, with errors scaling as the square
of the boulder diameter. A single boulder comparable to Yoshinodai on 25143
Itokawa, moved by as little as twice its own diameter, can alter the
magnitude of the torque by factors of several, and change the sign of
its spin component at all obliquities. Most of the total torque error produced
by multiple unresolved craters is contributed by the handful of largest
craters; but both large and small boulders contribute comparably to the total
boulder-induced error. A YORP torque prediction derived from groundbased data
can be expected to be in error by of order 100\% due to unresolved topography.
Small surface changes caused by slow spin-up or spin-down may have
significant stochastic effects on the spin evolution of small bodies. For
rotation periods between roughly 2 and 10 hours, these unpredictable changes
may reverse the sign of the YORP torque. Objects in this spin regime may 
random-walk up and down in spin rate before the rubble-pile limit is
exceeded and fissioning or loss of surface objects occurs. Similar
behavior may be expected at rotation rates approaching the limiting values for
tensile-strength dominated objects.

\end{abstract}


\end{frontmatter}


\section{Introduction}

Radiation-recoil effects are now understood to be major drivers in the
evolution of small bodies in the Solar System. The secular drift in orbital
elements produced by the Yarkovsky effect is a dominant mechanism in the
spread of asteroid families \citep{Bot01} and in the replenishment of the
near Earth object (NEO) population from the inner main belt \citep{Bot02}.
Torques produced by the YORP effect can compete with, and even
dominate, collisional and tidal torques on main belt asteroids and NEOs
smaller than a few km in diameter
\citep[and references therein]{Rub00,BotReview}.

Recent observational efforts have achieved direct detection of the influence
of the Yarkovsky and YORP mechanisms. Radar observations of 6489
Golevka show non-gravitational accelerations consistent with
Yarkovsky predictions \citep{Che03}. Photometric observations of 1862 Apollo
\citep{Kaa07} and 1620 Geographos \citep{Dur08b}, and photometric and radar
observations of (54509) 2000
PH$_5$ (subsequently named YORP) \citep{Low07,Tay07} indicate
accelerations in spin rate consistent with YORP models and not attributable
to tidal torques. However, the YORP models contain substantial systematic
uncertainties, because the objects' shapes and surface properties are not
sufficiently constrained by observations. In the case of 1862 Apollo, models
computed by \citet{Kaa04} with slightly different rotation pole positions
and shapes span a range of $\sim \pm 25\%$ around the measured acceleration.
For 54509 YORP, the models computed by \citet{Tay07}, using a range of
shapes and surface roughnesses, span a factor of $\sim 3$ in predicted
acceleration; and, in fact, they all systematically overestimate the observed
effect, by factors of 2 to 7.\footnote{Scaling down the models by these
factors was necessary to achieve the excellent fit in Fig.~2 of \citet{Low07},
as explained in the caption to that figure.}\ Considering the extensive
observational effort devoted to these objects, the ambiguity in the results
raises considerable concern as to whether YORP predictions derived from
groundbased light curve and/or radar data can ever be sufficiently precise
to permit quantitative tests of the theory.

This concern is only deepened by the history of YORP calculations made for
25143 Itokawa before and after the arrival of the Hayabusa spacecraft.
Using pre-encounter shape models derived from radar observations \citep{Ost04}
and photometric light curves \citep{Kaa03}, \citet{Vok04} predicted a
rotational acceleration of roughly half a part in $10^4$ per year, with an
estimated uncertainty of $\pm 30\%$. \citet{Sch07Itokawa} subsequently
recomputed the YORP prediction, first from a similar pre-encounter model
\citep{Gas06}, obtaining a result consistent with that of
\citet{Vok04}, and then from a sequence
of shape models of progressively higher resolution based on {\em in situ\/}
Hayabusa data. The first post-encounter shape model was found to imply a
rotational deceleration, meaning that pre-encounter models had been
unable to correctly predict even the sign of the effect. Subsequent models
incorporating the later Hayabusa data resulted in YORP accelerations that
did not converge as the resolution increased. \citet{Sch07Itokawa} conclude
that, even at sub-meter resolution, the essential geometry of the interaction
between the surface and the incoming and outgoing radiation is not adequately
represented. In the past year the situation has become even murkier. \citet{Sch08} show that the torque on 25143 Itokawa is very sensitive to the
position of the center of mass; and the predicted acceleration has still not
been detected \citep{Dur08a}.

The foregoing stories motivate this paper, which asks whether extreme
sensitivity to small-scale surface topography is a characteristic of the YORP
effect itself, or is limited to a comparatively small number of unfortunately
shaped objects. This question has pragmatic as well as predictive
consequences. First, dynamically significant small scale topography may be observationally unresolvable. I will show below that, indeed, neglecting
topographic features, such as moderate sized craters, that are
indistinguishable from the ground can lead to errors in YORP predictions of
order unity. As a result, shape models derived from groundbased data are
unlikely ever to yield precise predictions of YORP torques. Second, dynamically
significant small scale topography may,
in some situations, be unpredictably changeable. I will demonstrate that a
small displacement of a single large boulder can, in the right circumstances,
completely reverse the sign of the YORP torque. This implies that a YORP
cycle \citep{Rub00} can be interrupted in a major way by a comparatively
minor event, and suggests that the spin evolution of small NEOs may be more
stochastic than previously recognized.

The theoretical foundations and basic behavior of the YORP effect have been
analyzed extensively by analytic, semi-analytic, and numerical means
\citep[e.g.,][]{Sch07,Nes07,Nes08,Bre08,Mic08,Mys08a,Mys08b}. These works
have elucidated how asymmetries in the illuminated and radiating surface couple
to produce the now familiar behavior of the orbit-averaged torque
components as a function of obliquity \citep{Vok02}, and have provided
strategies for computing
torques for realistic asteroid models. This paper adopts a purely
numerical treatment, analyzing a large number of simulated objects
to yield statistically meaningful results in the nonlinear regime where
shadowing is important.

The remainder of the paper is laid out as follows. In Section \ref{s.method},
I describe the calculation of YORP torques for simulated asteroids.
I define a set of ``base objects'' on which the effect of
small-scale topography will be tested, and describe how topography
is added to the objects, in the form of Gaussian surface fluctuations,
craters, or boulders. In Section \ref{s.results}, I demonstrate that
small scale topography, at a level difficult or impossible to constrain
from groundbased observations, can typically be expected to alter YORP
torques by tens of percent or more. In particular, the magnitude, and even
the sign, of the torque can hinge on the location of a single large
crater or boulder. In Section \ref{s.discussion}, I discuss the implications
of these results, both for accurate predictions of the YORP effect and for
the YORP-driven evolution of small NEOs.

\section{Method\label{s.method}}
\subsection{Calculation of Torques\label{s.torque}}

Radiation interacting with the surface of a body can impart a torque through
the momentum it carries. It is conceptually helpful to split the interaction 
into three parts: (1) the momentum deposited on the surface by arriving
photons; (2) the recoil imparted to the surface by departing reflected photons;
and (3) the recoil imparted by departing reradiated photons. In this paper
I focus specifically on the third contribution, the reradiation torque.
The first contribution can be shown to cancel identically when averaged
around an orbit \citep{NV08}. The second is proportional to the albedo, and
becomes insignificant for a sufficiently dark surface; and, as \citet{Nes07}
point out, if the reflected and reradiated intensities are both assumed
isotropic, then the torque contributions are parallel.

I compute the torques on simulated objects using the TACO\footnote{``Thermophysical Asteroid Code, Obviously''}\ code, developed
by the author and students at Ohio University. The object's surface is
represented using a standard triangular tiling \citep{Lag96}. For all results 
presented here, 6,394 vertices and 12,784 tiles are used, for an average angular
resolution of approximately $7.5^\circ$ (3 tiles wide). For comparison, typical
shape models derived from asteroid light curve observations have 1,000 to 2,000
vertices and 2,000 to 4,000 tiles \citep[e.g.,][]{Dur07}. Models obtained
from radar observations and currently archived in the Planetary Data System
(PDS) have typically around 4,000 vertices and 8,000 tiles. Using
groundbased data alone, currently only 4 objects (1620
Geographos, 25143 Itokawa, 4179 Toutatis, and 52760) have been modeled at
resolutions equal to or finer than that used in this paper.

A horizon map (the maximum elevation $\psi_h$ of all visible parts of the
surface as a function of azimuth $\eta$, measured relative to local west)
is computed at the centroid of each tile when the surface is initially defined.
Shadowing is handled by requiring that the Sun not be blocked by other parts
of the surface for the tile to be illuminated. An illuminated tile
absorbs a flux $F_{\rm abs}$ given by
\beq
F_{\rm abs} = F \mu_0 \left[1 - r_{\rm hem}(\mu_0)\right],
\eeq
where $F$ is the incident flux, $\mu_0$ is the cosine of the angle between the
incident flux and the surface normal, and $r_{\rm hem}$ is the
directional-hemispheric reflectance or hemispheric albedo. TACO computes
$r_{\rm hem}$ using eq.~(42) of \citet{HapkeV},
assuming a single Henyey-Greenstein phase function. For definiteness I
adopt the mean Hapke parameters for S-class asteroids given by \citet{Hel89}: a
single-scattering albedo $w= 0.23$, and a phase function expanded to 6th order
in Legendre polynomials with a width parameter $\xi = -0.35$. The correction
for macroscopic surface roughness from \citet{HapkeIII} is applied with
a mean slope parameter ${\bar \theta}=20^\circ$; this has no bearing on the
absorbed flux but is relevant to the computation of light curves (section
\ref{s.results}). These reflectance parameters are assumed to be constant
over the surface. The opposition effect
\citep{HapkeV} is ignored, as is the illumination of other parts of the surface
by the reflected flux.

I neglect the effects of thermal conductivity, so that the
reradiated flux exactly balances the absorbed flux at all times. Each tile
is assumed to radiate isotropically; if there is no blockage by the local
horizon, the outgoing photons uniformly fill the upward-facing hemisphere,
and the recoil force on the tile is given by
\beq
f_{\rm rec} = 2 \pi \int_0^{\pi/2} {F_{\rm abs} A \over \pi c} \cos^2 \theta
	 \sin \theta d \theta = {2 F_{\rm abs} A \over 3 c},
\eeq
where $A$ is the tile area, $c$ is the speed of light, and $\theta$ is the
polar angle from the surface normal. In this case the recoil force is directed
inward, normal to the surface. If, instead, part of the sky
is blocked by the local horizon, some reradiated photons from the tile will
strike other parts of the object, and the tile will intercept some
photons reradiated by these other parts. Photons traded between
different parts of the surface cannot exert a torque on the body until they
are radiated into the clear sky. As a result, both the magnitude and direction
of the recoil force will be changed. I compute an approximate correction for
this effect, assuming that traded photons have no net effect on the
energy absorbed by each tile. For the moment, suppose that the tile's
local horizon were uniformly elevated an angle $\psi_h$ above the horizontal.
Then for the same radiated flux, the intensity would be increased by
a factor $\sec^2 \psi_h$, and the magnitude of the recoil force would be
given by
\beq
f_{\rm rec} = 2 \pi \int_0^{\pi/2-\psi_h} {F_{\rm abs} A
	\over \pi c \cos^2 \psi_h} \cos^2 \theta \sin \theta d \theta
	= {2F_{\rm abs} A \over 3 c} \left(1+{\sin^2 \psi_h \over 1 + \sin \psi_h}
	\right).
\eeq
The force would be increased, relative to the
flat-horizon case, by a factor $1 + \sin^2\psi_h/(1 + \sin \psi_h)$ because
the elevated horizon causes the outgoing radiation to be ``beamed'' into a
narrower solid angle. In practice, of course, the elevation of the horizon
is not uniform in azimuth. Therefore I replace $\psi_h$ with the mean
elevation, given by
\beq
\bar{\psi}_h = {1 \over N_h} \sum_{i=0}^{N_h-1} \psi_h(\eta_i),
\eeq
where $\eta_i$ is the azimuth of the $i$th point (out of $N_h=50$)
in the horizon map. The magnitude of the recoil force is then taken to be
amplified by a factor $1 + \sin^2\bar{\psi}_h/(1 + \sin \bar{\psi}_h)$
over the flat-horizon case.
The direction of $\vec{f}_{\rm rec}$ is taken to be
antiparallel to the clear-sky normal $\hat{N}_{\rm sky}$, which is computed
from a first-order Fourier expansion of the horizon map:
\beq
\hat{N}_{\rm sky} \equiv {\vec{n}_{\rm sky} \over n_{\rm sky}},
\eeq
where
\beq
\vec{n}_{\rm sky} = \hat{N}_{\rm surf}
- \hat{\epsilon}_{\rm n} \tan \left( {2 \over N_h} \sum_{i=0}^{N_h-1}
	\psi_h(\eta_i) \cos \eta_i \right)
- \hat{\epsilon}_{\rm w} \tan \left( {2 \over N_h} \sum_{i=0}^{N_h-1}
	\psi_h(\eta_i) \sin \eta_i \right),
\eeq
$\hat{N}_{\rm surf}$ is the surface normal, and $\hat{\epsilon}_{\rm n}$ and $
\hat{\epsilon}_{\rm w}$ are the unit vectors in the local north and west
directions, respectively.

The total
instantaneous torque is then simply the cross product
$\vec{r} \times \vec{f}_{\rm rec}$, where $\vec{r}$ is the vector from
the center of mass to the tile centroid, summed over tiles. The torque
is averaged over the spin and orbital phase, assuming the periods are
incommensurable. For computational simplicity, the orbit is taken to be
circular. For orbits with finite eccentricity $e$, the averaged torque can
be simply scaled by a factor $(1-e^2)^{-1/2}$ \citep{Sch07}.

\subsection{Base Objects\label{s.baseobjects}}

\begin{figure}[t]
\begin{center}
\includegraphics[width=3.0in]{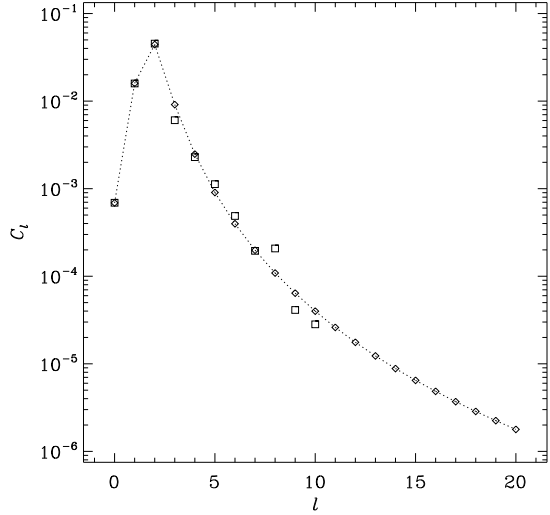}
\caption{
Coefficients of the Legendre expansion of the log-radius covariance function.
{\em Squares\/}: as tabulated by \protect{\citet{Mui98}}. {\em Diamonds and
dotted line\/}: from the adopted fit, Eq.~(\protect{\ref{e.fitcl}}).
\label{f.cl}
}
\end{center}
\end{figure}

I define a set of base objects, on which to measure the effects of smaller
scale topography, using the ``Gaussian random sphere'' formalism of
\citet{Mui98}. In this approach the distance to the surface from the origin
is given by
\beq\label{e.grsradius}
r(\theta,\phi) = \exp \left[ s(\theta,\phi) - {\beta^2 \over 2} \right],
\eeq
where $(\theta,\phi)$ are the usual spherical-coordinate angles and the
function $s(\theta,\phi)$ is defined by an expansion in spherical harmonics:
\beq\label{e.harmonicexpansion}
s(\theta,\phi) = \sum_{l=0}^\infty \sum_{m=-l}^l s_{lm} Y_{lm}(\theta,\phi).
\eeq
In this formulation, the mean radius of the body ${\bar r} \equiv 1$.
Since $s(\theta,\phi)$ is real, the coefficients $s_{lm}$ are constrained by
the relation
\beq
s_{l,-m} = (-1)^m s_{lm}^\ast.
\eeq
For nonnegative $m$, each $s_{lm}$ is a Gaussian random variable having zero 
expectation value and a variance given by
\begin{eqnarray}\label{e.grsvariance}
{\rm Var}[\Re(s_{lm})] &=& (1 + \delta_{m0}) {2 \pi \over 2l + 1} C_l;
\nonumber\\
{\rm Var}[\Im(s_{lm})] &=& (1 - \delta_{m0}) {2 \pi \over 2l + 1} C_l.
\end{eqnarray}
In eq.~(\ref{e.grsvariance}), $\delta_{ij}$ is the Kronecker delta symbol;
$C_l$ represents the coefficients of the Legendre expansion of the log-radius
covariance function, which also define the quantity
$\beta \equiv (\sum_l C_l)^{1/2}$ in eq.~(\ref{e.grsradius}). \citet{Mui98}
estimate the $C_l$ coefficients for $l \leq 10$ from shapes fitted
to light curve observations of 14 objects. For this work I adopt an analytic
fit to the results for their small-object subsample (4th column of
their Table 3), given by
\beq\label{e.fitcl}
C_l = 1.2 {(l^2 + 0.26)^2 \over (l^8 + 90.0)^{1.06}}.
\eeq
Fig.~\ref{f.cl} shows the coefficients from \citet{Mui98} and the fit. This
fit will be used below to extrapolate the covariance function out as far as
$l=40$. For $l \gg 1$, Eqs.~(\ref{e.grsvariance}) and (\ref{e.fitcl}) imply
$C_l \sim l^{-4.48}$, so that the
RMS amplitudes $s_{lm}$ decline as $l^{-2.74}$.

Each base object is a Gaussian random sphere with the expansion in 
Eq.~(\ref{e.harmonicexpansion}) truncated at a maximum order
$l_{\rm base}$. I adopt values of $l_{\rm base}=4,5,7,9,12,15,20$, and, for
each value, compute 6 different base objects with independent random
realizations of the $s_{lm}$ coefficients.

\subsection{Addition of Small-Scale Topography}

\begin{figure}[t]
\begin{center}
\includegraphics[width=3.5in]{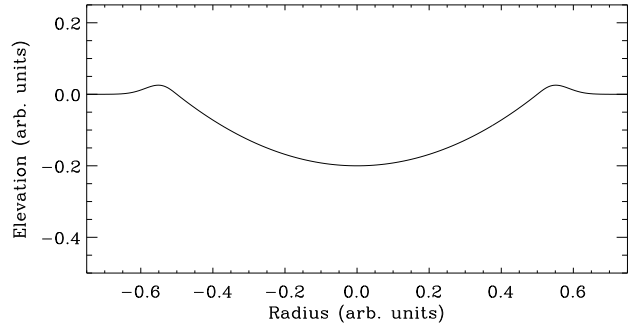}
\caption{
Crater profile according to Eq.~(\protect{\ref{e.craterprofile}}), with
depth to diameter ratio $q=0.2$ and rim width $\delta=D/20$.
\label{f.craterprofile}
}
\end{center}
\end{figure}

\begin{figure}[t]
\begin{center}
\includegraphics[width=4.5in]{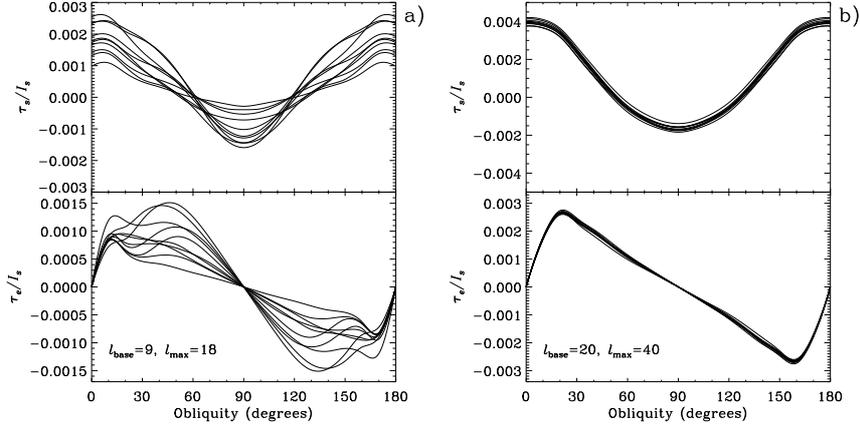}
\caption{
Typical variation in YORP torques caused by Gaussian topographic power added
at smaller scales to the base objects. ($a$) Base object with $l_{\rm base}=9$,
topography added to $l_{\rm max}=18$. Top- and bottom-most curves correspond
to the objects shown in Fig.~\protect{\ref{f.addobjects}}.
($b$) Base object with $l_{\rm base}=20$,
topography added to $l_{\rm max}=40$. Top panels show the spin component, and
bottom panels the obliquity component, of the torque, normalized by the moment
of inertia about the short (spin) axis, plotted against obliquity. Individual
curves correspond to ten different random realizations of the added topography.
\label{f.addexamples}
}
\end{center}
\end{figure}

The base objects are intended to represent asteroids whose shapes are
constrained by observations down to some minimum angular scale. Different
types of small-scale topographic structure can be added to the base objects.
In Section \ref{s.results} I measure the separate effects on YORP of each
of the following:
\begin{enumerate}
\item {\em Gaussian power at smaller scales.} Here the expansion in
Eq.~(\ref{e.harmonicexpansion}) is simply extended from $l_{\rm base}$ to
some higher maximum order $l_{\rm max}$. The $s_{lm}$ coefficients are identical
to those for the base object for $l \leq l_{\rm base}$, and chosen
randomly according to Eq.~(\ref{e.grsvariance}) for
$l_{\rm base} < l \leq l_{\rm max}$.
\item {\em Craters.} One or more circular craters are placed randomly on the
surface. The crater depth $z$ as a function of radius $r$ from its center is
given by
\beq\label{e.craterprofile}
z(r) = q D \left( 1 - 4 {r^2 \over D^2} \right) \times \cases{
1, $ if $ r \leq D/2;\cr
\exp\left[ -{(r - D/2)^2 \over 2 \delta^2} \right], $ if $ D/2 < r \leq D/2 + 3 
\delta.}
\eeq
In Eq.~(\ref{e.craterprofile}), $D$ is the crater diameter, $q$ is the
depth-to-diameter ratio, and $\delta$ is the width of the raised rim. I adopt
$q=0.2$ and $\delta = D/20$, based on NEAR-Shoemaker observations of 253
Mathilde \citep{Tho99} and 433 Eros \citep{Vev01,Tho02} and laboratory impact
experiments \citep{Nak02,Hou03}. The resulting profile is shown in
Fig.~\ref{f.craterprofile}. The vertical displacement $z$ is in the direction
of the mean surface normal of the area the crater will occupy. Prexisting
topography is first flattened to the level of the mean surface and then
overwritten by the crater. Multiple craters are added sequentially so that
overlapping features are reasonably realistic.
\item {\em Boulders.} One or more boulders can also be placed on the surface.
As a crude representation of a boulder, the tile vertices falling within a
given radius of the boulder center are raised, creating an approximately
round, mesa-like feature with jagged edges. As with craters, the existing
topography is first flattened so that the boulder's upper surface is
featureless.
\end{enumerate}
For each object, the center of mass and inertia tensor are calculated after the
small-scale topography is added,
assuming a homogeneous density $\rho$. The coordinate system is shifted and
rotated to align the origin with the center of mass and the coordinate axes
with the principal axes of the body. All objects are assumed to be rotating
about the shortest principal axis.

\section{Results\label{s.results}}
\subsection{Gaussian Power at Smaller Scales}\label{s.gaussian}

Any set of observations will be able to constrain the topographic power
down to some minimum length scale or maximum harmonic order. As a straightforward indicator of the effect of
neglected higher orders, I add power to each base object to a maximum
order $l_{\rm max} = 2 l_{\rm base}$, and measure the error in torque
incurred by ignoring topography down to half the observationally constrained
scale.

\begin{figure}[t]
\begin{center}
\includegraphics[width=4.0in]{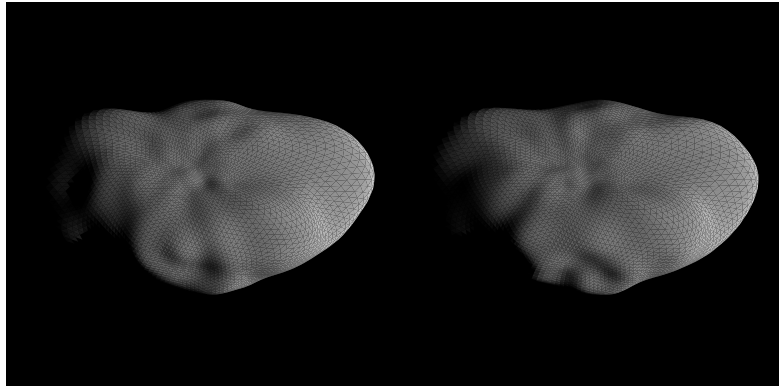}
\caption{
Identical views of the two $l_{\rm base}=9$ objects from
Fig.~\protect{\ref{f.addexamples}}a
with the ({\em left\/}) smallest and ({\em right\/}) largest torques.
Line of sight is in the plane containing the short and middle axes of the body,
$45^\circ$ from each,
and the illumination is from the right at $45^\circ$ phase angle.
\label{f.addobjects}
}
\end{center}
\end{figure}

Fig.~\ref{f.addexamples} shows typical results for two of the 42 base
objects, one with $l_{\rm base}=9$ and one with $l_{\rm base}=20$, corresponding
to minimum angular scales of $40^\circ$ and $18^\circ$, respectively. 
The curves show the averaged torque components affecting the spin rate
($\tau_s$) and the obliquity ($\tau_\epsilon$), normalized by the principal
moment of inertia about the short axis $I_s$, plotted against
obliquity.\footnote{Numerical results for normalized torques are given in
units of $1.52 \times 10^{-15} (\rho / 2{\,\rm g\,cm^{-3}})^{-1}
({\bar r} / 1{\rm \, km})^{-2} (a/1{\,\rm AU})^{-2} (1-e^2)^{-1/2}
{\rm\, s}^{-2}$, where
$\rho$ and ${\bar r}$ are the constant density and
mean radius of the body, and $a$ and $e$ are the orbital semimajor axis
and eccentricity.}\ Different curves
indicate the 10 random realizations of
the added topography. Clearly, knowing the topography only to order
$l_{\rm base}=9$ (Fig.~\ref{f.addexamples}a) is not sufficient to determine the
torque to even order-unity accuracy. The range in $\tau_s$ spans nearly a
factor of 3 at obliquity $\epsilon=0^\circ$, and more than a factor of 5 at
$\epsilon=90^\circ$. The location of the ``Slivan states'' \citep{Sli02},
which require
$\tau_s=0$, varies over $15^\circ$; and the obliquity torque $\tau_\epsilon$
in the vicinity of the Slivan states varies
over a factor of 5. In contrast, the $l_{\rm base}=20$ objects
(Fig.~\ref{f.addexamples}b), to which power has
been added to $l_{\rm max}=40$, have largely concordant torques. Evidently,
at sufficiently small angular scales, the neglect of Gaussian power at still
smaller scales is not too serious, as long as it follows a simple extrapolation
of the covariance coefficients.

As a measure of how wrong a YORP prediction is likely to be due to neglect
of the smaller scale topography, I compute, at each value of obliquity, the
standard deviation in each torque component over the 10 trials, then average
over obliquity; this quantity is normalized to the mean absolute value of
that component, also averaged over obliquity. The result is an
obliquity-independent measure of the expected relative error.
For the $l_{\rm base}=9$ case in Fig.~\ref{f.addexamples}a, the expected
relative errors are 38\% in $\tau_s$ and 32\% in $\tau_\epsilon$.
Keep in mind that
these are $1\sigma$ errors, implying 95\% confidence bands (for a Gaussian
distribution) with full widths of 152\% and 128\%, consistent
with the order-unity variation seen in the figure. For the $l_{\rm base}=20$
case, the relative $1\sigma$ errors are 7\% and 3\% in $\tau_s$ and
$\tau_\epsilon$
respectively. I will use this same approach to quantify the effects of craters
and boulders below.

The most remarkable thing about the results in Fig.~\ref{f.addexamples}a is
that the individual objects, despite differing by factors of a few in their
YORP torques, appear nearly identical. Fig.~\ref{f.addobjects} shows
a representative view of the two objects in Fig.~\ref{f.addexamples}a
having the most discordant torques. To assess whether these objects would
be distinguishable photometrically, I compute optical light curves
over a grid of 40 rotation pole directions and 10 phase angles from
$0^\circ$ to $90^\circ$. I find that the RMS photometric difference between
these two objects is $<0.05$ magnitudes at all phase angles. The {\em maximum\/}
difference is $<0.1$ mag at phase angles $\leq 50^\circ$, and is $>0.2$ mag
only at phase angles of $75^\circ$ or higher, making it unlikely that these
objects would be distinguished by groundbased observations.

\begin{figure}[t]
\begin{center}
\includegraphics[width=4.0in]{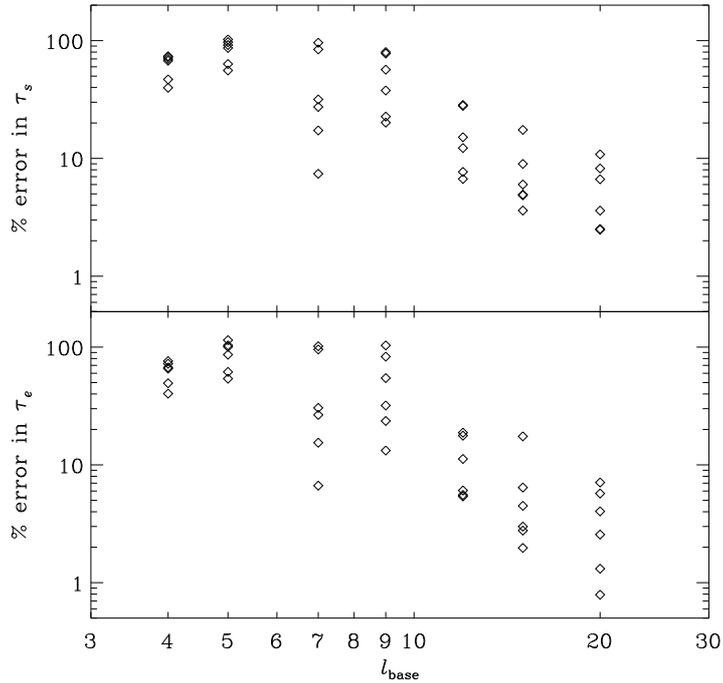}
\caption{
Expected relative $1\sigma$ error in the YORP torque caused by neglecting
Gaussian topographic power at $l_{\rm base} < l \leq 2 l_{\rm base}$. 
{\em Top\/}: error in the spin component; {\em bottom\/}: error in the
obliquity component. The relative error is the standard deviation in each
component over 10 random trials, averaged over obliquity, and normalized to
the mean absolute value of that component. Multiple symbols correspond to 6 different base objects at each $l_{\rm base}$.
\label{f.adderrors}
}
\end{center}
\end{figure}

\begin{figure}[t]
\begin{center}
\includegraphics[width=4.0in]{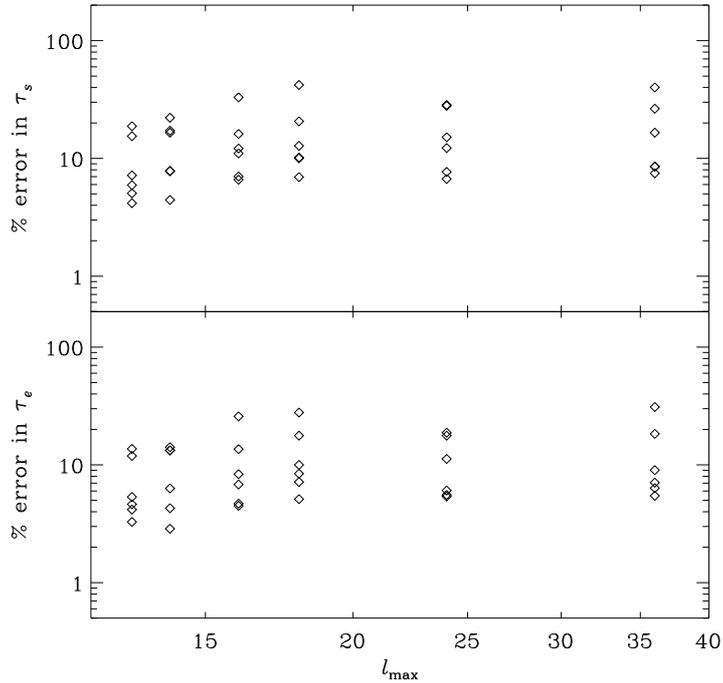}
\caption{
Expected relative $1\sigma$ error, as in Fig.~\protect{\ref{f.adderrors}},
for the $l_{\rm base}=12$ base objects with Gaussian power added to
$l_{\rm max} = 13, 14, 16, 18, 24, 36$. There is no strong dependence on
$l_{\rm max}$ for $l_{\rm max}>18$,
indicating that orders just above the cutoff $l_{\rm base}$ are most important.
\label{f.addmaxerror}
}
\end{center}
\end{figure}

In Fig.~\ref{f.adderrors}, I show the relative 1-sigma errors for
all of the base objects. The scatter at each $l_{\rm base}$
reflects the base objects' different YORP susceptibilities. One
can easily see that order-unity errors in the torque should be expected for
$l_{\rm base} \leq 10$. To be reasonably confident of errors under 10\%, the
surface must be known to at least $l_{\rm base} = 20$.
Owing to the steepness of the extrapolated covariance function,
most of the change in the torque is actually produced by orders close to
$l_{\rm base}$. Fig.~\ref{f.addmaxerror} shows the relative errors for the
$l_{\rm base}=12$ objects with power added to $l_{\rm max}=13,14,16,18,24,$
and $36$. There is no discernible increase in the error for $l_{\rm max}>18$,
and the single order $l=13$ seems to be responsible for roughly half of the
asymptotic value.

\subsection{Craters}\label{s.craters}

Actual small-scale topography is likely to be non-Gaussian. Impact
craters are one example of Poisson-distributed features not well described
by a truncated spherical-harmonic expansion. The crater size distribution
has been estimated for 951 Gaspra \citep{Cha96b,Gre94}, 243 Ida \citep{Bel94,Cha96a}, 253 Mathilde \citep{Cha99}, and 433 Eros \citep{Cha02}.
To lowest order, the cumulative distribution of craters with diameters
larger than $D$, $N(D)$, is reasonably well approximated by a power law,
$N(D) \sim D^{-2}$ \citep{OBr06}. Though there are significant deviations in the 
logarithmic slope for different objects, the known
distributions all describe a steeply diminishing number of large craters. The
largest craters have diameters ranging from $0.35$ to $1.2$
times the bodies' mean radii \citep{Tho99}.

\begin{figure}[t]
\begin{center}
\includegraphics[width=4.0in]{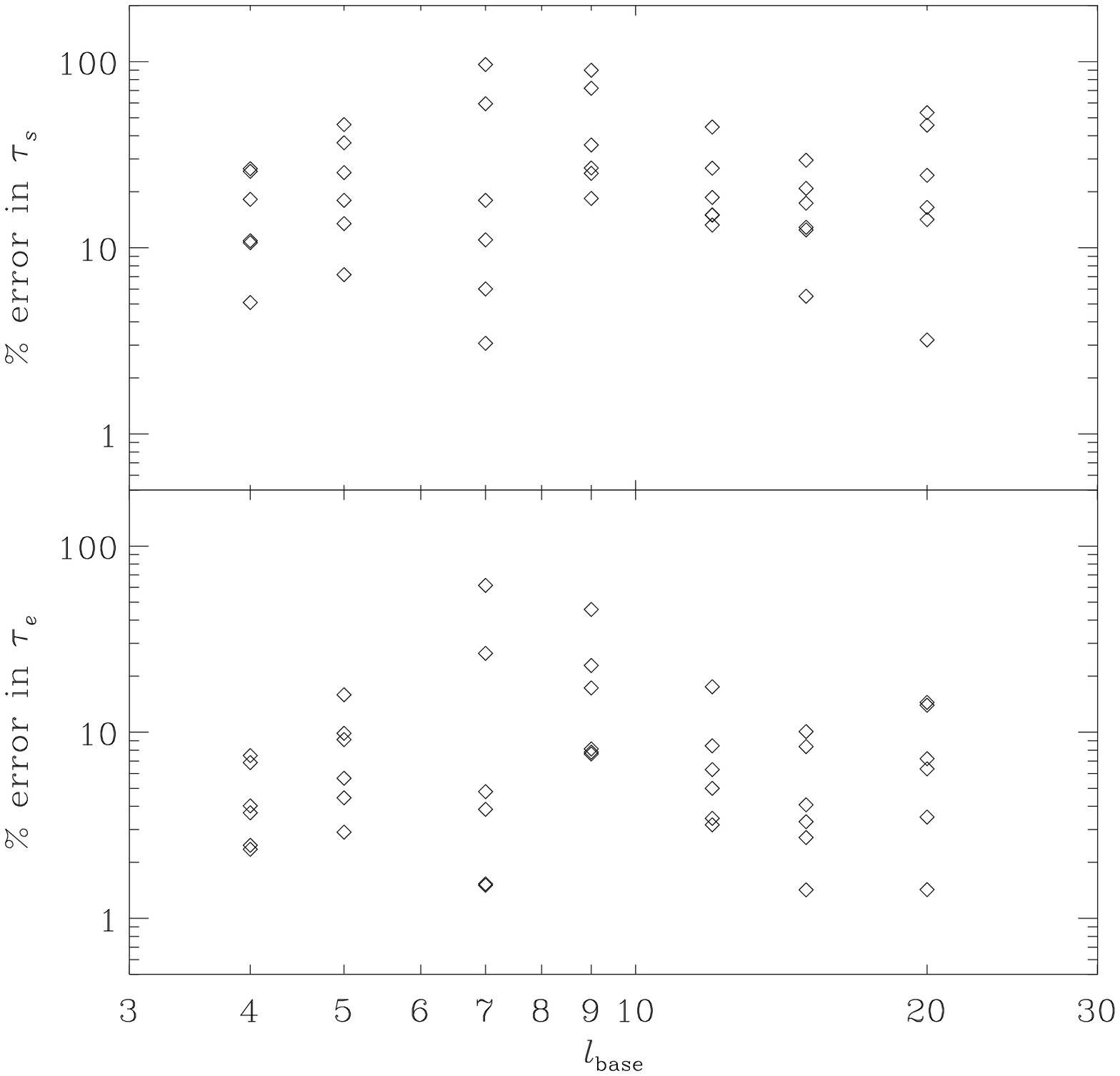}
\caption{
Expected relative $1\sigma$ standard deviation in the YORP torque, as in
Fig.~\protect{\ref{f.adderrors}}, caused by a single randomly placed crater
with diameter $D=0.3$ added to different base objects.
\label{f.craterrors}
}
\end{center}
\end{figure}

I return to the original base objects, removing the small-scale Gaussian power
added in Section \ref{s.gaussian}, in order to measure separately the effect
of craters on the YORP torque components. I start by randomly placing a single,
modest-sized crater of diameter $D=0.3$
(in units of the body mean radius) on the base objects. Fig.~\ref{f.craterrors}
shows the resulting relative $1\sigma$ errors, calculated as in
Sec.~\ref{s.gaussian} with ten random crater placements for each object. A
single crater can clearly alter the spin and obliquity torques by tens of
percent, even though the crater itself, including the raised rim, occupies only
$\sim 0.5\%$ of the total surface. The effect increases with $l_{\rm base}$
at values $l_{\rm base} \lesssim 9$. A similar, but less pronounced, increase
is also seen in the Gaussian case (Fig.~\ref{f.adderrors}). The single crater
actually produces less error for $l_{\rm base} \lesssim 9$ than does small-scale
Gaussian power added to the same base object. But the effect of the crater for
$l_{\rm base}>10$ is larger
than the Gaussian case and, to within the statistical errors, independent of
the base order. This may be due either to the declining amplitudes $C_l$ at
higher $l$ or to the fact that $l=9$ corresponds roughly to the scale of the
crater itself. Evidently the crater couples to other mesoscale surface
features of roughly its own size more effectively than to the large-scale
structure at smaller $l$.

\begin{figure}[t]
\begin{center}
\includegraphics[width=4.5in]{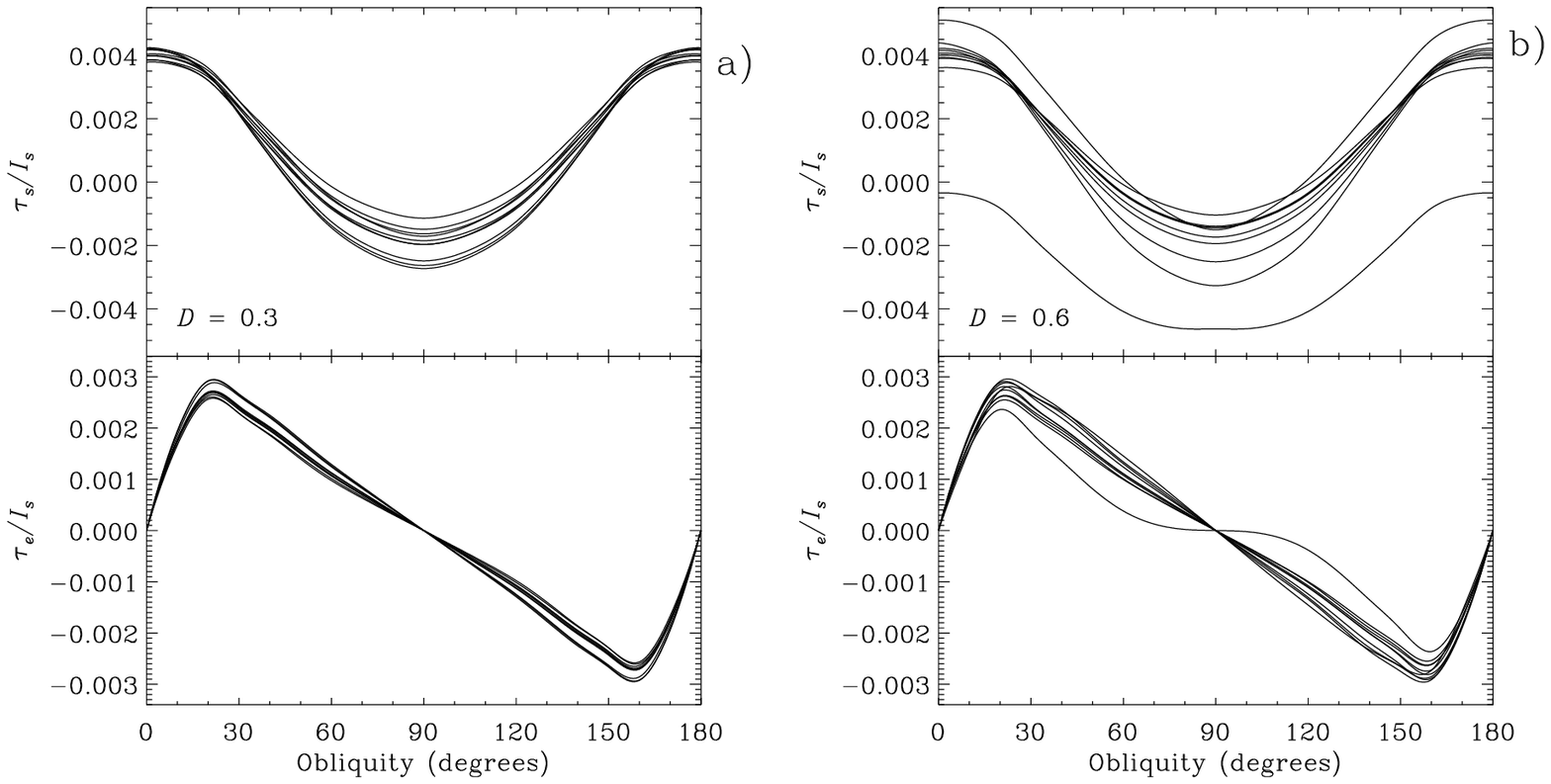}
\caption{
Typical variation in YORP torques caused by single craters 
of diameter ($a$) $D=0.3$, ($b$) $D=0.6$, randomly placed
on the same base object used in Fig.~\protect{\ref{f.addexamples}}b.
Top- and bottom-most curves correspond
to the objects shown in Fig.~\protect{\ref{f.cratobjects}}.
\label{f.cratexamples}
}
\end{center}
\end{figure}

\begin{figure}[t]
\begin{center}
\includegraphics[width=4.0in]{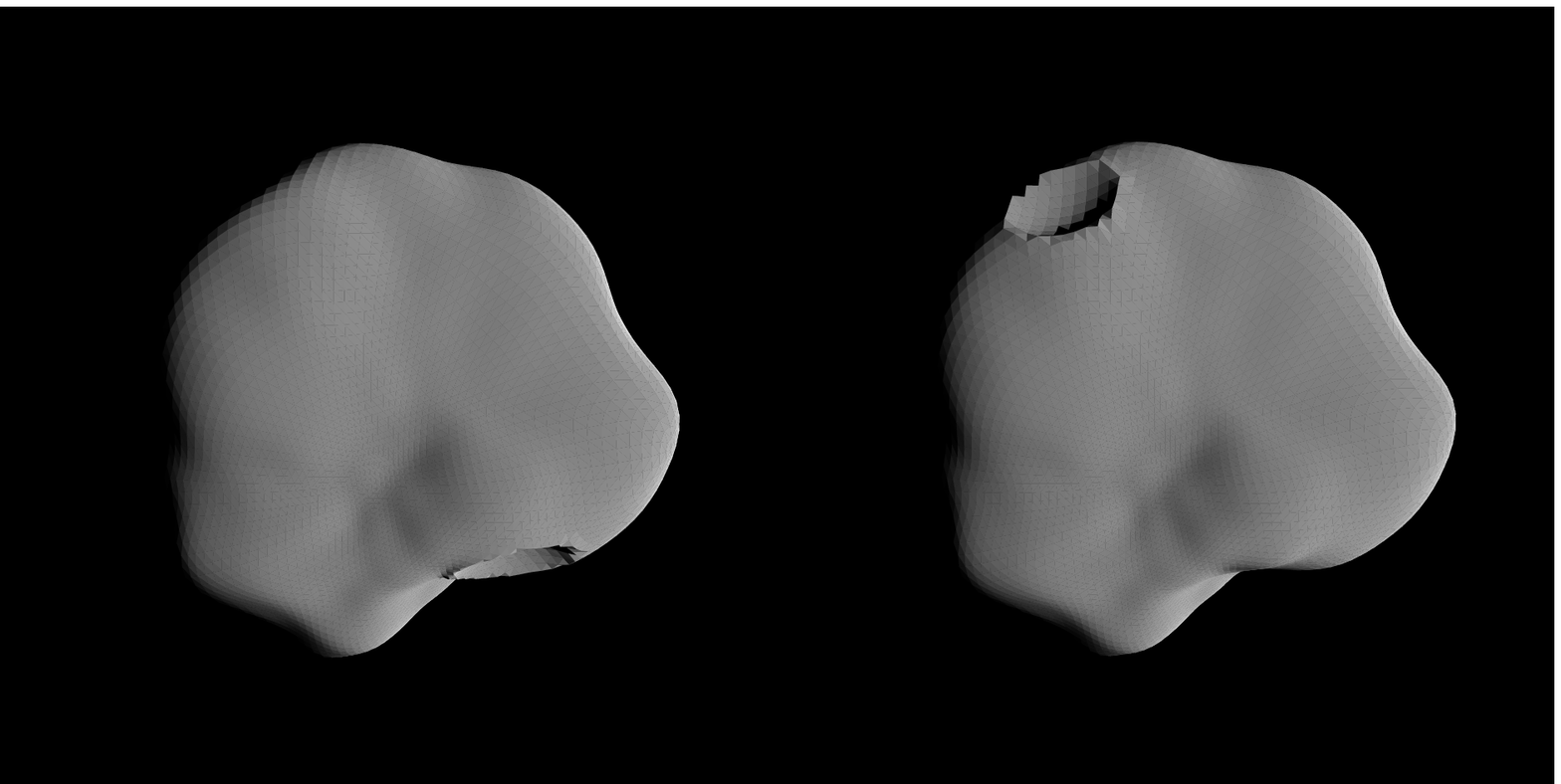}
\caption{
Identical views, along the rotation axis, of the two objects having the
({\em left\/}) smallest and ({\em right\/}) largest torques in
Fig.~\protect{\ref{f.cratexamples}}b.
\label{f.cratobjects}
}
\end{center}
\end{figure}

Fig.~\ref{f.cratexamples}a shows the normalized torque as a function of
obliquity for ten random placements of single craters on the same
$l_{\rm base}=20$ base object used in Fig.~\ref{f.addexamples}b. The effect
of the crater is about twice that of the added Gaussian topography: the
relative errors in $\tau_s$ and $\tau_\epsilon$ are 14\% and 6\%, respectively,
which also represent median results over all base objects. Fig.~\ref{f.cratexamples}b
shows the effect of doubling the crater size to $D=0.6$ on the same base object.
The relative errors increase to 56\% and 14\%. Comparing the extreme examples
(the highest and lowest curves), the spin components differ by a factor of
$4.4$ at $\epsilon=90^\circ$ and actually have opposite signs
at $\epsilon=0^\circ$.
Fig.~\ref{f.cratobjects} shows these two most discordant objects, which turn
out to be cratered at almost diametrically opposite positions.

Cratering a base object moves the center of mass and slightly reorients the
principal axes. To determine whether these adjustments contribute to the
change in torque, I recompute the torques for 60 of the cratered and bouldered
(see Section \ref{s.boulders}) objects, without shifting to the new center of
mass or reorienting to the new principal axes. For either craters or boulders,
the results differ, on average,
by only a few percent, indicating that it is really the surface topography
that is responsible for the large expected torque errors.

\begin{figure}[t]
\begin{center}
\includegraphics[width=4.0in]{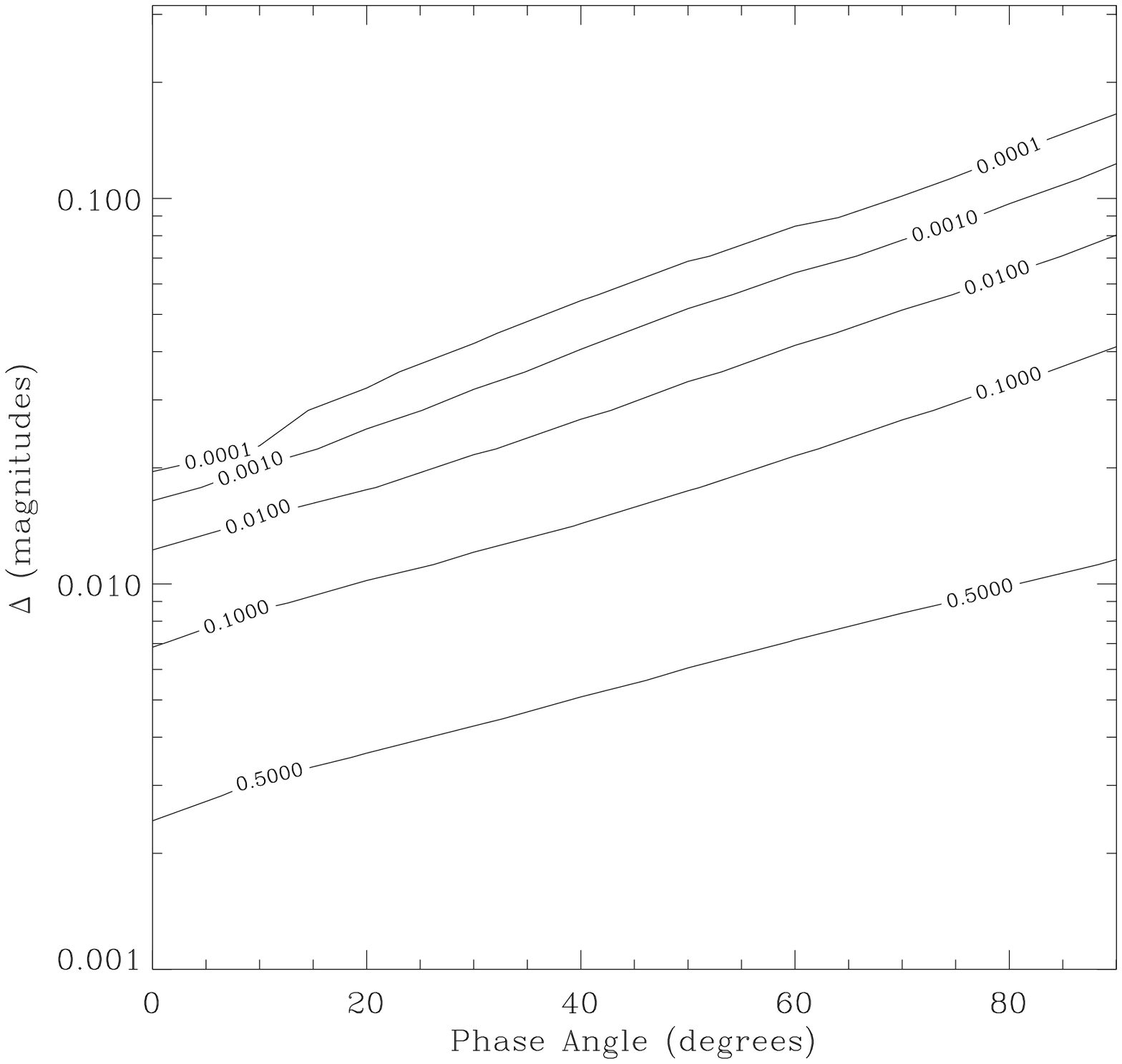}
\caption{
Probability contours for observing a photometric difference $\Delta$ or larger,
as a function of phase,
between pairs of objects that differ only in the location of a single crater
of diameter $D=0.6$. Results shown are averaged over the 6 base objects with
$l_{\rm base}=20$.
\label{f.cratphotometry}
}
\end{center}
\end{figure}

The objects, even those with single large ($D=0.6$) craters, are so
photometrically
similar that it is unlikely that they would be distinguishable in current
light curve programs. To assess this possibility, I take all 60 of the
$l_{\rm base}=20$
cratered objects and compute light curves over the same grid of 40 orientations
and 10 phase angles used in section \ref{s.gaussian}. I then calculate the
probability that two objects, differing only in crater location, would be
observed at a given moment to have a magnitude difference $> \Delta$. The
result, as a function of phase angle, is shown in Fig.~\ref{f.cratphotometry}.
Note that, unless substantial data are obtained at phase angles $>70^\circ$,
one expects that no more than 1\% of the total data will differ by more than
$0.05$ mag, and no more than $0.01$\% by more than $0.1$ mag. In practice,
observations at phase angles above $45^\circ$ are rare in light curve programs.

\begin{figure}[t]
\begin{center}
\includegraphics[width=4.0in]{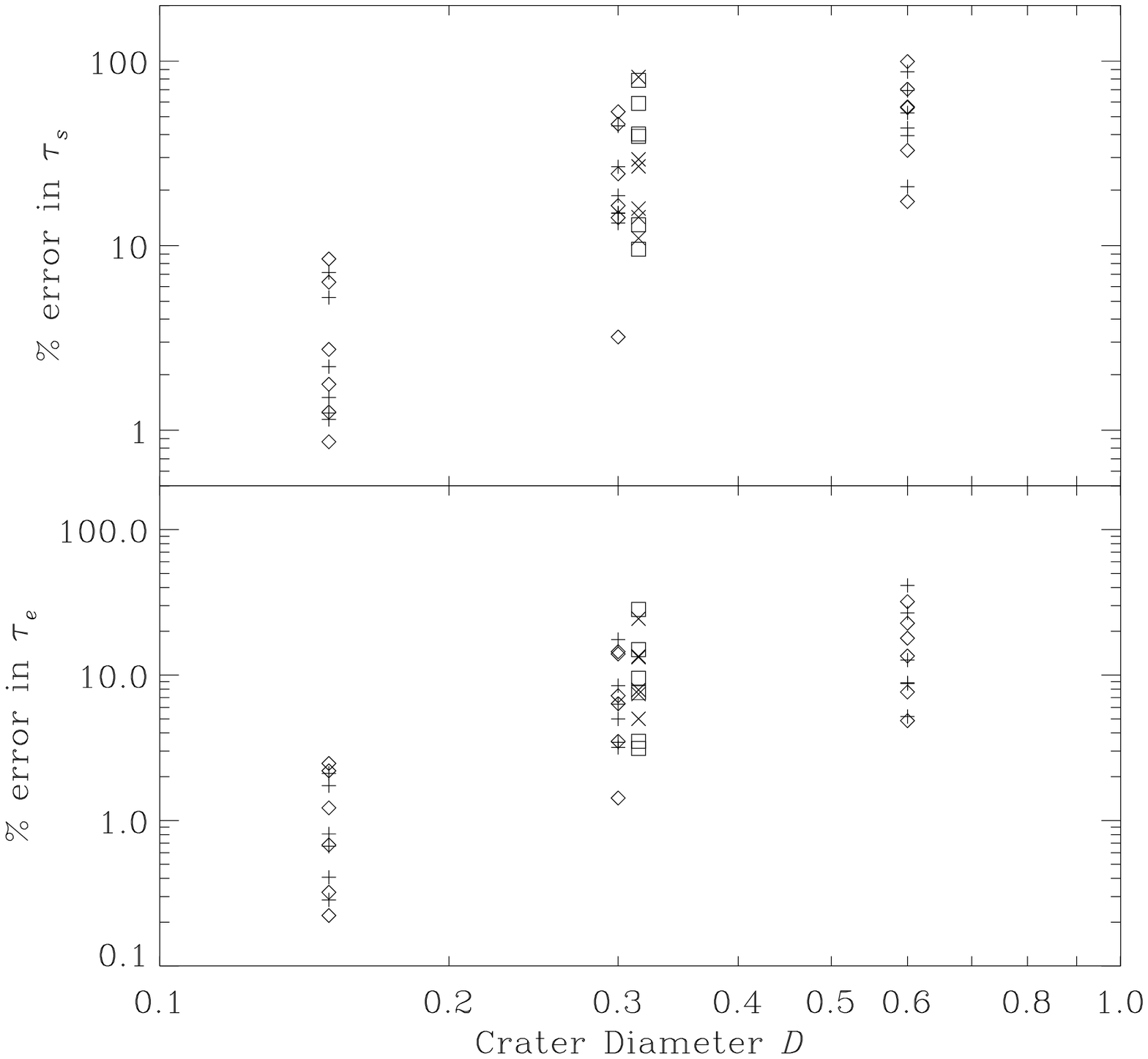}
\caption{
Expected relative error in torque, as in Fig.~\protect{\ref{f.craterrors}},
caused by single craters randomly placed on base objects with $l_{\rm base}=12$ ({\em plus signs\/}) and $l_{\rm base}=20$ ({\em diamonds\/}), as a function
of crater diameter $D$. {\em Crosses\/} and {\em squares\/} (offset
horizontally) show results for 4 randomly placed craters each of diameter
$D=0.3$, on objects with $l_{\rm base}=12,20$ respectively.
\label{f.cratsizeerrors}
}
\end{center}
\end{figure}

Fig.~\ref{f.cratsizeerrors} shows the scaling of the relative torque errors
with crater size. The diamonds and plus signs show results for single-cratered
objects with $l_{\rm base}=12$ and 20, respectively. One might naively
expect the torque error to scale with the crater area ($\sim D^2$); and
Fig.~\ref{f.cratsizeerrors} shows that
this is not a bad description for $D>0.3$. However, at smaller diameters the
scaling is much steeper, with the relative error roughly $\propto D^3$.
This nonlinear response may be attributable to shadowing, although the
same steepening does not occur for boulders (Section \ref{s.boulders} below).
The squares and crosses in Fig.~\ref{f.cratsizeerrors} (slightly offset for
clarity) indicate the results for objects with four $D=0.3$ craters each. Since
the changes in torque produced by each added crater are random and uncorrelated,
one should expect the total change to scale as the square root of the number
of craters. This expectation is confirmed by the factor 2 increase in the
relative errors over the single-crater case. The implications of these
scaling laws for the expected accuracy of YORP predictions on realsitically
cratered objects are discussed in section \ref{s.discussion} below.

\subsection{Boulders}\label{s.boulders}

Boulders are a second example of Poisson-distributed topography. The surface
of 25143 Itokawa has been found to be dominated by boulders, with a cumulative
size distribution well represented by $N(D) \propto D^{-3}$ over a factor
10 in diameter; the largest boulder, Yoshinodai, is fully one-tenth the size of
its parent body \citep{Sai06}. For
433 Eros, \citet{Cha02} obtain a result consistent with $N(D) \propto D^{-4}$.
These size distributions are substantially steeper than those for craters;
thus one can expect the cumulative effect of moderate sized boulders to
be more relatively important than that of moderate sized craters.

I proceed with the same strategy to assess the influence of boulders on YORP
torques. I add
a single mesa-shaped boulder, with diameter $D=0.3$ (in units of the body
mean radius) and height $h=D/2$, at
random locations to the base objects. The base objects have mean
longest diameters of approximately $3.1$, so the size of these boulders
relative to their parent bodies approximately matches that of Yoshinodai. The height-to-diameter ratio
corresponds to typical values measured for boulders on 433 Eros \citep{Tho02}.

\begin{figure}[t]
\begin{center}
\includegraphics[width=4.0in]{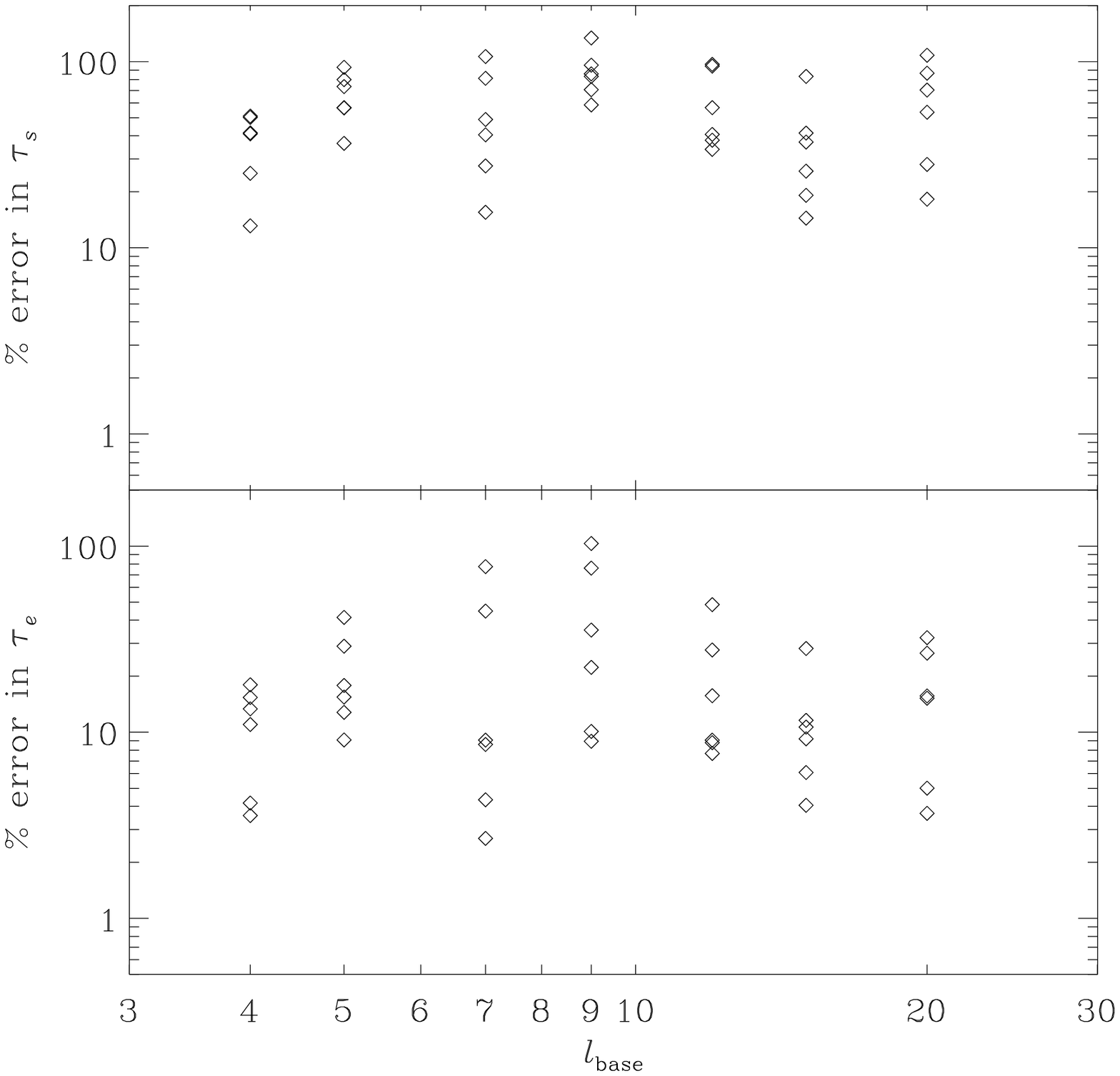}
\caption{
Expected relative $1\sigma$ standard deviation in the YORP torque, as in
Fig.~\protect{\ref{f.adderrors}}, caused by a single randomly placed boulder
with diameter $D=0.3$ added to different base objects.
\label{f.boulerrors}
}
\end{center}
\end{figure}

\begin{figure}[t]
\begin{center}
\includegraphics[width=4.0in]{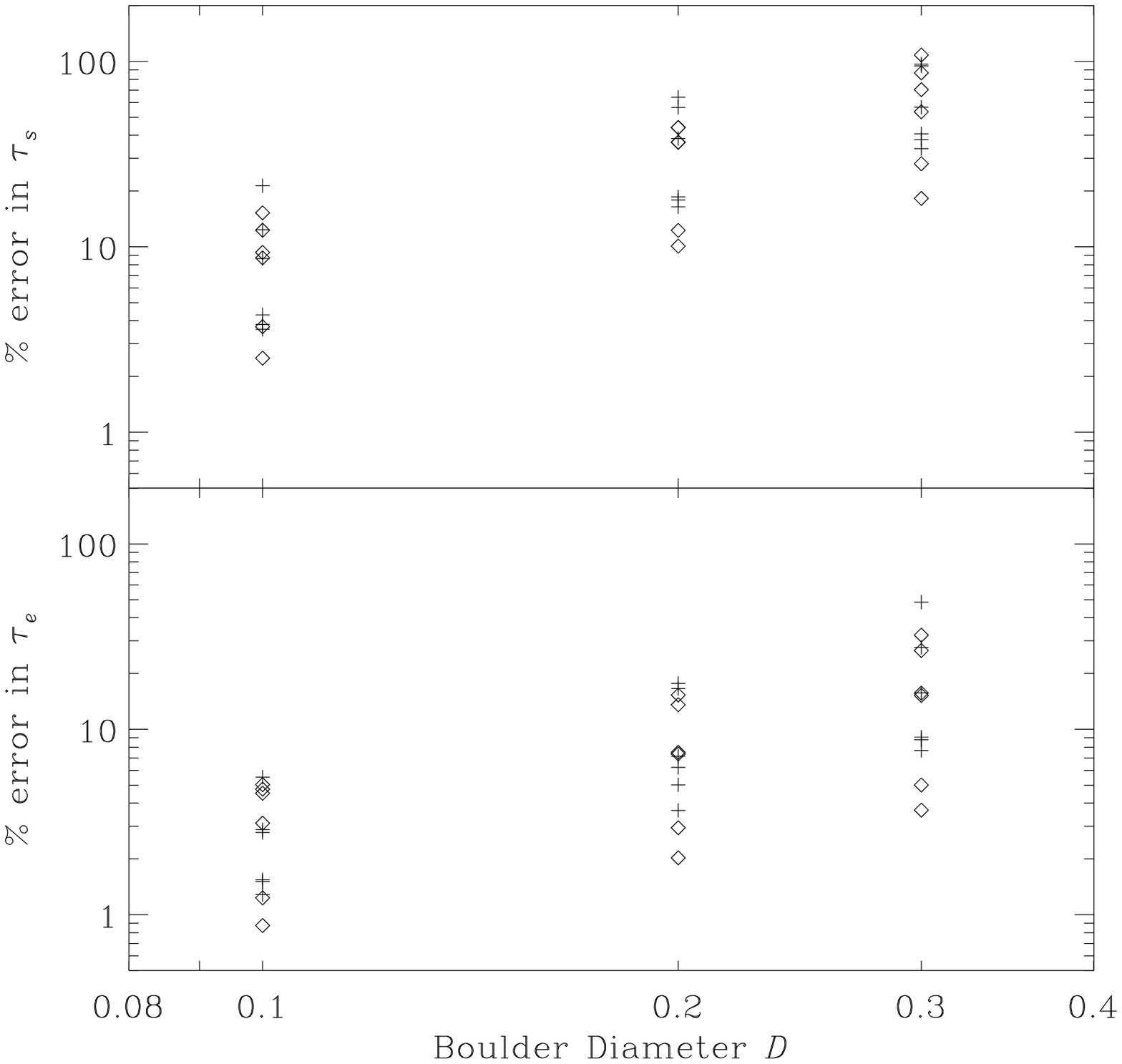}
\caption{
Expected relative error in torque, as in Fig.~\protect{\ref{f.boulerrors}},
caused by single boulders, as a function of boulder diameter $D$. Symbols
have the same meaning as in Fig.~\protect{\ref{f.cratsizeerrors}}.
\label{f.boulsizeerrors}
}
\end{center}
\end{figure}

Fig.~\ref{f.boulerrors} shows the expected relative error in the torque
components caused by the single large boulder. The effect is
quite dramatic: typical $1\sigma$ errors are 10\% to 100\% in the spin
component, and 5\% to 50\% in the obliquity component. This is roughly a
factor 3 larger than the effect of craters of the same diameter, though
the difference may be attributable to the simple fact that boulders can be
proportionally
taller than craters are deep (here, by a factor of $2.5$). As in the case of
craters, there is a weak increase in the mean error for values of $l_{\rm base}
< 10$, but little difference is seen at higher order.
The scaling of the errors with boulder size is shown in
Fig.~\ref{f.boulsizeerrors}. The boulders are all similarly shaped, with
$h = D/2$. The relative error scales quite cleanly as $D^2$, i.e., proportional
to the boulder surface area.

\begin{figure}[t]
\begin{center}
\includegraphics[width=4.5in]{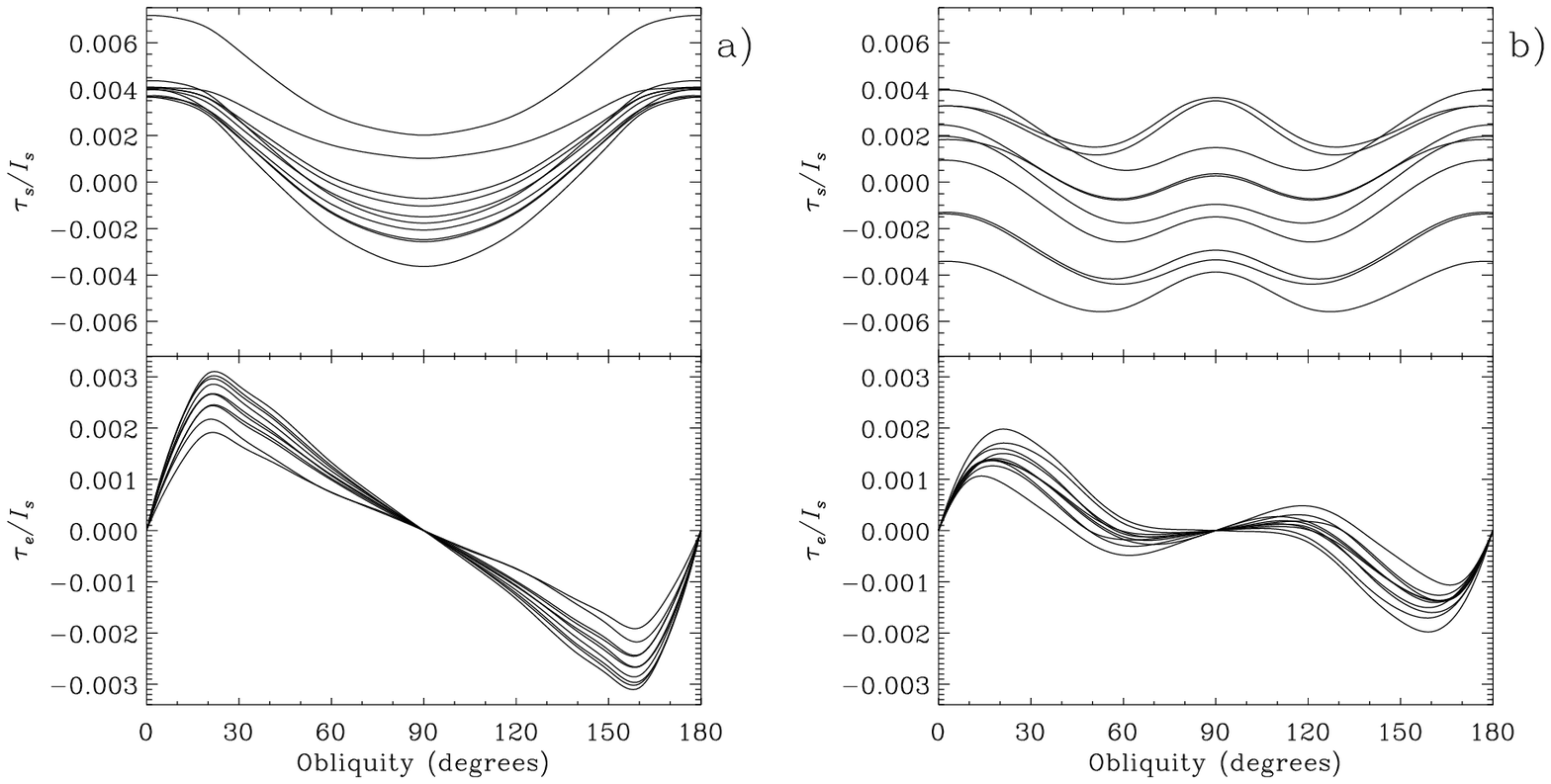}
\caption{
Typical variation in YORP torques caused by single boulders 
of diameter $D=0.3$, randomly placed on ($a$) the same base object used
in Figs.~\protect{\ref{f.addexamples}}b and \protect{\ref{f.cratexamples}},
and ($b$) a second object with $l_{\rm base}=20$, showing more extreme
variation. Top- and bottom-most curves correspond
to the objects shown in Fig.~\protect{\ref{f.boulobjects}}.
Note in ($b$) the possibility of reversing the sign of the spin
torque at all obliquities simply by repositioning one boulder.
\label{f.boulexamples}
}
\end{center}
\end{figure}

\begin{figure}[t]
\begin{center}
\includegraphics[width=4.0in]{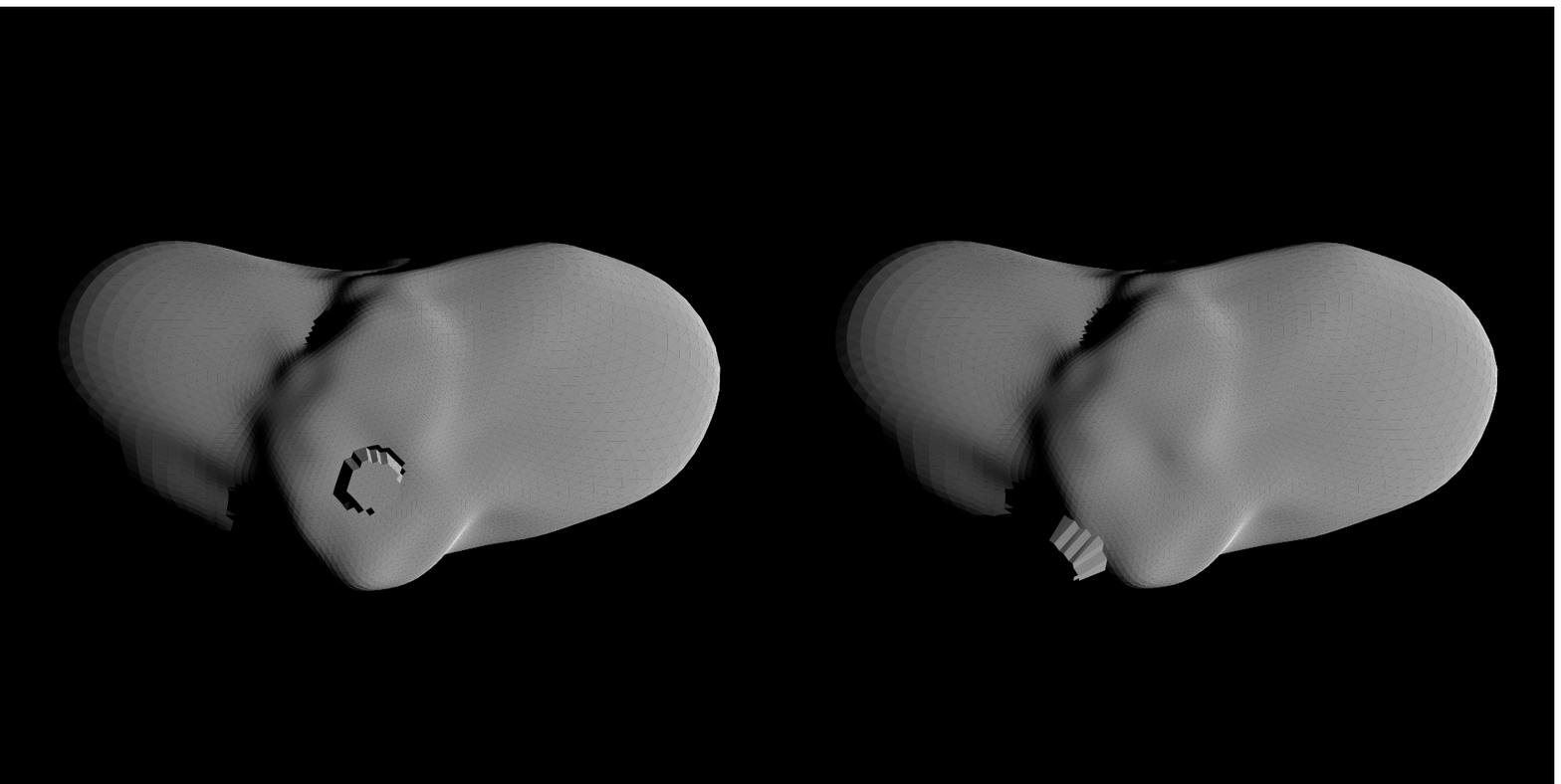}
\caption{
Identical views of the two objects from Fig.~\protect{\ref{f.boulexamples}}b
with the ({\em left\/}) most positive and ({\em right\/}) most negative
spin torques. Line of sight is in the plane containing the short and middle
axes of the body, $45^\circ$ from each, and the illumination is from the right at $30^\circ$ phase angle. Moving the boulder by only twice its own diameter
reverses the sign of the spin torque.
\label{f.boulobjects}
\label{lastfig}			
}
\end{center}
\end{figure}

Examples of the torque components as a function of obliquity are shown in
Fig.~\ref{f.boulexamples}a, for 10 random $D=0.3$ boulder placements on the same
$l_{\rm base}=20$ object used in Figs.~\ref{f.addexamples}b and
\ref{f.cratexamples}. This case again represents a typical amount of
topographically induced variation. Boulders have a tendency
to shift the curves for $\tau_s/I_s$ up and down while altering the overall
amplitude of those for $\tau_\epsilon/I_s$. Note that the sign of the spin
torque
can be flipped by the location of a single boulder, even for this unexceptional
object. A more extreme example is seen in Fig.~\ref{f.boulexamples}b, which
shows the same curves for one of the other five $l_{\rm base}=20$ objects. Here
the sign of the spin torque at all obliquities can be reversed by the location
of a single Yoshinodai-sized boulder. The two most discordant objects are shown
in Fig.~\ref{f.boulobjects}. The boulder location differs by barely twice its
own diameter; yet the object on the left will inexorably spin up while the
object on the right will spin down.

\section{Discussion}\label{s.discussion}

The above results indicate that sensitivity to small-scale topography is an
inherent characteristic of the YORP effect. Since YORP operates by virtue of
the non-cancellation of many independent torque components that vary across
the surface, over the rotation of the body, and around the orbit, this
conclusion is perhaps not surprising; however, it does impose strong
limits on the predictability of YORP torques, and has important consequences
for the spin evolution of small bodies.

If the surface is describable by a spherical harmonic expansion whose
coefficients are Gaussian-distributed, the topography must be known
accurately at least to harmonic order $l=10$ to allow a YORP prediction
of even the correct sign. To achieve 10\% accuracy requires the surface
to be known at least to $l=20$, if the topography diminishes with $l$ in
a manner consistent with a simple extrapolation of the \citet{Mui98}
covariance coefficients (Eq.~[\ref{e.fitcl}]). Few objects can be modeled
reliably at this resolution. Furthermore, actual small scale topography is
not Gaussian and not small in amplitude. The above results show that
Poisson-distributed features such as craters and boulders, which produce
substantial shadowing and which are not well revealed by groundbased
observations, can have significant effects. A single large feature
can alter YORP torques by tens of percent or more.

I found above that the relative torque error scales as $\sigma \sim D^{\beta}$,
where $\beta\approx 2$ for boulders and  $2 \lesssim \beta \lesssim 3$ for
craters. Using these numerical results and the observed
size distributions, one can estimate the relative influence of large and
small craters and boulders. The observed cumulative
counts are reasonably well represented by $N(D) \sim D^{-\alpha}$, with
$3 \lesssim \alpha \lesssim 4$ for boulders and $1 \lesssim \alpha
\lesssim 2$ for craters.\footnote{As before, diameters are implicitly in
units of the body mean radius.}\ Assuming the effects of individual boulders or
craters to be uncorrelated, one expects in either case that the variance
in torque produced by features with diameters between $D_{\rm min}$ and
$D_{\rm max}$ will scale according to
\beq\label{e.cratbouleffects}
\sigma^2 \propto \int_{D_{\rm min}}^{D_{\rm max}}{dN \over dD}D^{2 \beta}dD
\propto D_{\rm max}^{2\beta-\alpha} - D_{\rm min}^{2\beta-\alpha},
         \qquad (\alpha \ne 2 \beta).
\eeq

For cratered objects, Eq.~(\ref{e.cratbouleffects}) implies that at least
75\% of the total variance (87\% of the expected $\sigma$) is produced by
the 4 biggest craters. Fig.~\ref{f.cratsizeerrors} shows that the expected
error from the random location of the single largest crater is already
approaching 100\% at $D_{\rm max}=0.6$, close to the observed maximum diameter
on Eros
but rather modest compared to the values of $0.89$ and $1.2$ for Ida and
Mathilde, respectively \citep{Tho99}. These very large craters could possibly
give rise to noticeable features in high phase-angle light curves; however,
light curve inversion performs poorly in recovering surface concavities
because the inversion problem is non-unique \citep{Kaa92,Kaa01a}. The convex,
pre-encounter model of the {\em Rosetta\/} target 2867 Steins \citep{Lam08}
resembles the {\em in situ\/} images,\footnote{\tt
{http://www.esa.int/SPECIALS/Rosetta/SEMNMYO4KKF\_0.html}}\ but
by construction misses several large craters, including one with
an apparent diameter $D\approx 0.9$. A non-convex inversion probably would
not have revealed this crater, as the light curve
data are limited to phase angles $< 42^\circ$. Convex
models, in general, may be useful for determining overall surface reflectance
and thermal properties, and may give accurate predictions for Yarkovsky accelerations. But a YORP prediction based on a convex model is likely to be
wrong by of order 100\%.

Boulders, because of their steeper size distribution, behave differently:
{\em at most\/} half of the total variance is produced by boulders larger
than half the maximum size. In fact, if the distribution
is as steep as on Eros ($\alpha = 4$), then the variance is
formally divergent as $D \to 0$. Of course, distributions $dN/dD \sim D^{-1}$
or steeper are unphysical in this limit; but the scaling indicates that
the YORP torque can be affected as much by the surface distribution of small
boulders as by that of large ones. The results of section \ref{s.results} also
show that single boulders may have significantly greater effect on the torque
than craters of the same size; a single Yoshinodai-sized object can easily
alter the spin torque by tens of percent, and in some cases change its sign.
The clear implication is that the surface boulder distribution can be
critical in determining the actual consequences of YORP.

Thus, one should expect uncertainties of order unity in predictions
of the YORP effect made from models derived from remote observations, just from
the unconstrained nature of the small-scale surface topography. The true
situation is actually even worse, since I have neglected other issues, such
as variations in albedo, surface roughness, and internal density, which
probably have comparable effects. Indeed, \citet{Sch08}
find that shifting the center of mass of Itokawa by only 15 m can reverse
the sign of the spin torque. At present, there have been 4 attempts
to predict the YORP acceleration from groundbased data and compare with direct
measurements. In one case (1682 Apollo), the correct magnitude and sign were
obtained, with an uncertainty of $\pm 25\%$; in a second (54509 YORP), models
gave the correct sign but the wrong magnitude by a factor of 2 to 7. In
the third (25143 Itokawa), pre-Hayabusa models evidently
obtained the wrong sign (compared to later models), but the predicted
acceleration has still not been detected. And in the fourth (1620 Geographos),
torques computed for different convex shape models bracketed the observed
value, but varied over a factor of 5. This situation is about what we should
expect. Generally speaking, a YORP prediction computed from a groundbased
model will be accurate only to within factors of order unity, and will
have no better than an 80\% chance of having the correct sign.

The sensitivity of reradiation torques to boulders also implies that
there may be substantial stochasticity in the spin evolution of
small bodies under the influence of YORP. In the standard picture of the
YORP cycle \citep{Rub00}, an object can gradually spin up, simultaneously
evolving in obliquity, as determined by the now familiar ``YORP curves''
(e.g., Figs.~\ref{f.addexamples}, \ref{f.cratexamples}, and
\ref{f.boulexamples}). In most cases, spin-up continues only until 
reaching an orientation regime where $\tau_s < 0$.
The object then spins down, asymptotically approaching the stable fixed
point in obliquity. Objects with substantial power in the higher YORP
orders \citep{Nes07,Nes08} can sometimes have stable fixed points at
obliquities where $\tau_s > 0$; examples of this behavior are seen in
Figs.~\ref{f.cratexamples}b and \ref{f.boulexamples}b, along with cases
where $\tau_s > 0$ at all obliquities. These objects would presumably
spin up indefinitely. But a new possibility raised
here is that a seemingly minor topographic change, such as the movement
of a boulder, can stochastically shift the evolution onto a new set of
YORP curves. One would expect such events to occur well before catastrophic
fissioning or even unbinding of surface material, as the local effective
gravity vectors change with the accelerating spin. Calculations for
contact-binary configurations \citep{Sch07b} and numerical experiments on
rubble piles \citep{Wal08,Har09} confirm that surface changes occur at rotation
rates slower, by factors of several, than the nominal centrifugal limit for
self-gravitating spheres. The latter corresponds to the observed
minimum rotation periods of large asteroids, at approximately 2 hours. Thus
a broad range of spin periods, from 2 to perhaps 10 hours, may
correspond to a ``scattering zone'', in which small
topographical changes stochastically reverse the YORP torques.
An object being spun up by YORP may scatter multiple times
through motion or loss of surface boulders, random-walking up and down in
spin rate, before finally crossing the actual spin limit and either fissioning
or shedding all loose objects. It is extremely interesting that the
observed distribution of NEOs in the diameter-rotation period plane
\citep[e.g., Fig.~1 of][Statler et al.\ 2009, in preparation]{Hol07}
shows the sharp cutoff at 2 hours becoming notably indistinct at diameters
$\lesssim 1\,{\rm km}$, where YORP is expected to become dominant.
Similarly, small, fast-rotating objects approaching the higher spin limit imposed by tensile strength \citep{Hol07}
may wander up and down in spin rate, driven not by
catastrophic fragmentation, but merely by a subtle rearrangement of the
surface that reverses the YORP effect.

The author is grateful to David Riethmiller, Desiree Cotto-Figueroa, and
Kyle Uckert for their assistance in the development of TACO; to Nalin
Samarasinha, Dan Scheeres, Mangala Sharma, and Desiree Cotto-Figueroa for
extensive comments on the manuscript; and to the referees, David
Vokrouhlick\'y and David Rubincam, for important suggestions that improved
the paper. This work has made use of NASA's Planetary Data System and
Astrophysics Data System Bibliographic Services.

\label{lastpage}


\bibliographystyle{elsarticle-harv}
\bibliography{asteroids}

\end{document}